**Giant Magnetoelectric coupling in Single Phase $Pb(Zr_{0.20}Ti_{0.80})_{0.70}Pd_{0.30}O_{3-\delta}$ Multiferroics**

Shalini Kumari[1], Dhiren K. Pradhan[1], Nora Ortega[1], Kallol Pradhan[1], Christopher DeVreugd[2], Gopalan Srinivasan[2], Ashok Kumar[3], J. F. Scott[4], and Ram S. Katiyar[1,*]

[1]Department of Physics and Institute for Functional Nanomaterials, University of Puerto Rico, San Juan, PR 00931-3334, USA.

[2]Physics Department, Oakland University, Rochester, Michigan 48309-4401, USA.

[3]CSIR-National Physical Laboratory, Dr. K. S. Krishnan Marg, New Delhi 110012, India

[4]Department of Chemistry and Department of Physics, University of St. Andrews, St. Andrews KY16 ST, United Kingdom.

**Abstract:**

During the last fifteen years the multiferroic (MF) research communities have been searching for an alternative room temperature MF material with large magnetoelectric (ME) coupling for possible applications in high density electronic components, and low heat dissipation memory and logic devices.[1,2,3,4,5,6,7,8] In the past few years, we have investigated several multi-component systems such as $Pb(Zr,Ti)O_3$(PZT) - $Pb(Fe,Ta/Nb/W)O_3$, and related family members, which have shown better ME effects compared to bismuth ferrite.[9,10,11,12,13,14,15] In continuation to our search for larger ME effect, we have studied $Pb(Zr_{0.20}Ti_{0.80})_{0.70}Pd_{0.30}O_{3-\delta}$ (PZTP30) system with an unusually large (30%) palladium occupancy in B site of PZT. This material exhibited a giant ME coupling coefficient ~0.36 mV/cm.Oe. Interestingly, this is the first time any room temperature single phase compound that showed ME trends, and magnitude similar to those in the well established mechanical strain-mediated ferroelectric and ferromagnetic composites; the latter ones are already in the commercial stage as nT/pT magnetic field sensors due to their large ME values. Our new system is a simple tetragonal crystal structure with space group *P4mn* as probed by X-ray diffraction (XRD), and Raman studies. A neutral Pd atom (Kr $4d^{10}$) has square planar complex, which gives zero magnetic moment (μ=0) and diamagnetism, but when it is in $Pd^{+2}$ or $Pd^{+4}$ ionic states, it provides large magnetic moment with outer cell configuration $4d^{8}5s^{0}$ or $4d^{6}5s^{0}$ with unpaired electrons in its d-shell. It has also been proven that electric control can change diamagnetic Pd to ferromagnetic, magnetism in clusters of Pd atoms, and large magnetization in multilayer structures with Pt and Fe.[16,17,18,19] The presence of Pd in PZTP30 has been confirmed by XPS and XRF studies and assigned with related binding energies of $Pd^{+2}$ and $Pd^{+4}$ ions as 336.37 eV, 342.9 eV, and 337.53 eV, 343.43 eV, respectively, which may be the origin of room temperature magnetism in Pd substituted PZT ceramics. A sharp first order ferroelectric phase transition was observed at ~569 K (+/-5 K) that is confirmed from dielectric, Raman, and thermal analysis. Both ferromagnetic and ferroelectric orderings with large ME coupling were found above room temperature, a significant step forward in the development of single phase ME material with enhanced functionalities.

[*]Author to whom correspondence to be addressed. Electronic mail: rkatiyar@hpcf.upr.edu (Prof. Ram S. Katiyar).

## Introduction

Large room-temperature magnetoelectric effects in single-phase material is one of the aims of scientists working worldwide in the area of multiferroics, similar to the desire for room-temperature superconductors. These systems require the presence of simultaneous ferroic order parameters with strong ME coupling for an increased number of logic states. The basic physics and coupling mechanisms between spin and polarization in crystals, especially for single phase

ME systems, is not well studied yet for the next generation of logic and memory elements.[20,21,22,23,24] Due to the natural chemical incompatibility between magnetism and ferroelectricity in oxide perovskites, only a few single-phase multiferroic oxides exist with sufficiently large magnitude of polarization and magnetization for real device applications. Some of the well known potential multiferroic materials are as follows: $BiFeO_3$, $YMnO_3$, $Pb(Fe_{0.5}Nb_{0.5})O_3$, $Pb(Fe_{0.5}Ta_{0.5})O_3$, $Pb(Fe_{0.67}W_{0.33})O_3$ $TbMnO_3$ etc.[8,15,25,26,27] The most well known room-temperature lead-free single-phase multiferroic is $BiFeO_3$ (BFO), having both ferroelectric ( $T_C = 1143$ K) and antiferromagnetic ( $T_N = 643$ K) phase transition above room temperature; however, still it is not suitable for real practical device applications due to high leakage current and small ME coupling coefficient.[28,29,30,31] Multiferroics possess ferroic order parameters and their cross coupling at cryogenic temperatures (either ferroelectric and/or magnetic transitions), among them the magnitude and directions of magnetic and ferroelectric orders often occur largely independent of each other; and as a result, the magnetoelectric coupling tends to be small.[32,33,34] The low operational temperatures (ferroelectric/magnetic), high leakage current, and/or weak ME coupling of most of the single-phase compounds have motivated researchers to continue the search of novel room temperature magnetoelectric multiferroics with larger ME coupling coefficients.

An alternate option is ferroelectric and ferromagnetic composites/heterostructures which have greater design flexibility for ME devices by combining ferroelectric (FE) and ferromagnetic (FM) phases together; among them a heterostructure with FE/FM phases is the most popular design, due to immense potential for high density logic states.[35] Single phase magnetoelectric materials with large ME coupling coefficients in ultrathin heterostructured films are important for tunnel junction-based devices, which can provide a higher degree of logic states under combined electric and magnetic control. These factors drive the endeavors for discovering new room temperature single-phase multiferroics with giant ME coupling, beyond $BiFeO_3$. Several materials have recently been discovered, such as $Pb(Zr_{0.53}Ti_{0.47})_{0.60}(Fe_{0.5}Ta_{0.5})_{0.40}O_3$ (PZTFT), the Bi-based aurivillius oxides, $GaFeO_3$, and $LuFeO_3$.[12,36,37,38,39]

Therefore, the discovery of an alternative room temperature single phase multiferroic material is highly exciting. In this report an attempt has been made to realize multiferroism and ME coupling at room temperature. $PbZr_xTi_{1-x}O_3$, which is known to be one of the best ferroelectric materials nature whereas, palladium is a transition metal without any magnetic properties.[40,41] However, we have realized multiferroicity (ferroelectricity and magnetism) with strong intrinsic ME coupling at room temperature in the palladium-substituted PZT.

**Methods**

Polycrystalline powder of $Pb(Zr_{0.20}Ti_{0.80})_{0.70}Pd_{0.30}O_{3-\delta}$ (PZTP30) were synthesized using a conventional solid-state reaction route from a stoichiometric mixture of highly pure (>99.99%) reagents PbO, $ZrO_2$, $TiO_2$, and PdO powders from Alfa Aesar. Mechanical ball milling of stoichiometric amounts of all ingredients was carried out in methanol, followed by calcination in air at 1150 $^oC$ for 8h, using a carbolite furnace (HTF1700) with heating rate of 5$^oC$/min. The synthesized phase pure powder was pressed into pellets (d=13 mm) at a uniaxialforce of 5 tons and later sintered at 1200 $^oC$ for 8h. Phase formations of the pellets were evaluated by X-ray diffractometer (Rigaku Ultima III) equipped with a Cu$K_\alpha$ radiation source operating in the Bragg–Brentano geometry at 40 kV and 40 mA in a slow-scan mode 0.2$^o$/min. The Rietveld structure refinement was carried out using FullProf Suite Software. Scanning electron microscopy (SEM) images were recorded at 5000X magnification with help of a JEOL JSM-6480LV system operated with an accelerating voltage of 20 kV to study the surface morphology. The composition and valance states of the fabricated pellets were confirmed via X-ray fluorescence spectroscopy (XRF) and high-resolution X-ray photoemission spectroscopy (XPS), correspondingly. Structural analysis was made via temperature-dependent Raman spectroscopy in addition with XRD using a T64000 spectrometer operating in backscattering configuration and in subtractive mode. About 8 mW of continuous wave power from a Coherent argon ion laser (Innova 90–5) at 5145 Å was focused to a small spot size of ~ 2 $\mu m^2$. A liquid nitrogen-cooled CCD device collected the low and high temperature Raman scattered signal through an 80X objective in vacuum from 83 K to 900 K in steps of 25 K using Linkam TP93 and TMS94 temperature controllers, and liquid nitrogen pump cooling module. The flat surface of the sintered pellets was polished with fine emery paper, and then top and bottom electrodes were made by coating high purity silver paint followed by heating at 200 $^oC$ for 2 h in air for better conduction and adhesion. The dielectric measurements were performed under vacuum ($10^{-6}$ Torr) using an HP4294A impedance analyzer. Thermal control was achieved in the range of 200 – 700 K using a variable temperature micro-probe system equipped with a programmable temperature controller (MMR Technologies, Inc.). Ferroelectric properties were measured using Radiant RT 6000 High Voltage System after poling the sample under a voltage of 1000 V for 6 h using DC Power supply (TREK, Inc., Model: 677A) at room temperature. Low temperature magnetic properties of the PZTP30 samples were measured using a Quantum design PPMS DynaCool in a wide range of temperature 25 - 400 K. The room temperature magnetoelectric measurements were carried out with a homemade ME-set up using a magnet with varying field of up to ±3kOe with lock-in amplifier and reference ac magnetic field, using a Helmholtz coil.[42]

The Rietveld refinement of high resolution XRD data provides straightforward and precise structural information. A detailed XRD studies have been performed on PZTP30 pellets used for magnetic, and ME measurements with assumption of A-site occupancy by Pb and B-site by Zr, Ti, and Pd with oxygen at the corner of $BO_6$ octahedral position. The Rietveld refinement of the diffraction patterns was performed by considering the tetragonal *P4mm* symmetry.[43] The experimental and Rietveld simulated XRD patterns of PZTP30 bulk samples are shown in Figure 1(a). The results demonstrated excellent fit, confirming pure tetragonal phase formation of the material belonging to the space group *P4mm.* The sharp Bragg peaks were assigned to their Miller indices appearance of any extra reflection peaks that would be indicative of secondary phases and peaks from lead-deficient pyrochlore phases. We refined many parameters, such as background, zero shift, specimen displacement, atomic positions, thermal factors, scale factor, lattice parameters, FWHM, and shape parameters. Pseudo–Voigt description of profile shape was taken into account as a profile set-up for Rietveld refinement. The difference between the measured spectrum and the refined one is very small, and the reliability is ensured by the refinement parameters. All atoms were fixed to their site occupancies, as their variation did not appreciably affect the refinement results. During the refinement process it was observed that reliability factors improve further when anisotropic thermal parameters were taken into account compared with the isotropic thermal parameters of the individual atoms. The crystal structure parameters, and reliability factors obtained after XRD refinement are listed in Table S1. The calculated tetragonality ratio was c/a= 1.047, which indicates suitability for large polarization. Using the obtained unit cell parameters and atomic positions, the three dimensional sketch of tetragonal PZTP30 unit cell projected along c axis has been simulated, as shown in Figure 1(b), which confirms that Pd was incorporated into the crystalline lattice of the PZT; hence the appearance of tetragonality in PZTP30 is consistent with PZT(20/80).[43,44,45] The bond lengths obtained after XRD refinement has been compiled in Table S2. The overall microstructure of PZTP30 pellets showed well defined densely packed grains with average size ranging between ~ 3–10 $\mu$m surrounded by distinct grain boundaries as shown in Fig. S4. The presence of different shapes and grain sizes with neck to neck compaction revealed that the grain growth process was almost completed during the sintering process. The average XRF data matched with the initial elemental compositions taken for this study within the experimental limitations as shown in Table S3.

For conclusive evidence of existence of all elements with their valance states, PZTP30 pellets were examined by high-resolution X-ray photoelectron spectroscopy (XPS). Figures 2 and S5 show the core level high resolution XPS spectra of Pb 4*f*, Zr 3*d*, Ti 2*p*, Pd 3*d,* and O 1*s*. The peak positions in the XPS spectra were referenced to C 1*s* peak at 284.8 eV. The observed binding energy positions of Pb are Pb $4f_{7/2}$ =138.2 eV, and Pb $4f_{5/2}$ = 143 eV respectively. The high resolution XPS spectrum of Zr 3*d*, and Ti 2*p* splits up into two components due to spin-orbit effects. The spin-orbit doublets of Zr, and Ti were observed at following binding energies: Zr $3d_{5/2}$ = 181.3 eV, and Zr $3d_{3/2}$ = 183.7 eV; Ti $2p_{3/2}$ = 457.8 eV, and Ti $2p_{1/2}$ = 463.5 eV. These energy values confirm the valance states of $Zr^{4+}$, and $Ti^{4+}$ ions without any trace of $Ti^{3+}$ ions (which could in principle be responsible for some magnetism). Note that $Pd^{4+}$ is an excellent fit in the B-site since it and sixfold-coordinated $Ti^{4+}$ have the same (0.061-0.062 nm) ionic radii. $Pd^{2+}$ is too small to fit well in the $Pb^{2+}$ site. The Pd $3d_{5/2}$ and Pd $3d_{3/2}$ doublets were deconvoluted into $Pd^{2+}$ at binding energy 336.37 eV, 342.9 eV, and $Pd^{4+}$ at binding energy 337.53 eV, 343.43 eV respectively in PZTP30 ceramics (see Figure 2(a)).[46] Its effective area (occupied by $Pd^{4+}$ ions) is quite large compared to $Pd^{2+}$ ions (ratio:4/3), which is the origin of magnetism in Pd-substituted PZT ceramics. The O1*s* can be deconvoluted into two peaks with binding energies of 531.1 eV and 529.3 eV with shallower binding energy peak belonging to the lattice oxygen and the deeper binding energy attributed to the surface adsorbed oxygen (see Figure 2(b).[47] The binding energies for all the individual elements match the standard value.[48] Moreover, the presence of all elements in PZTP30 bulk was confirmed clearly through XPS measurement along with XRF measurement.

Temperature and frequency dependent dielectric studies were carried out over a wide range of temperatures and frequencies to understand the ferroelectric to paraelectric phase transition behavior, and dielectric dispersion characteristic. The order and nature of the phase transition allows us to understand the domain dynamics above and below the Curie temperature ($T_c$). The variation of relative dielectric permittivity ($\varepsilon_r$) and loss tangent (tan δ) as a function of temperature at different frequencies for PZTP30 is shown in Figure 3(a) and its inset, respectively. The dielectric permittivity decreases with increase in frequency for PZTP30, which is a signature of polar dielectrics. Here $\varepsilon_r$ increases with increase in temperature, reaches a maximum, and then decreases. This observed sharp anomaly at ~569 K (+/- 5K) represents the ferroelectric-paraelectric transition temperature ($T_C$). This further increases above ~600 K due to thermally activated charge carriers. The temperature dependent tan δ (inset of Figure 3(a)) also shows an anomaly just below $T_c$ above 1 kHz probe frequencies. At high temperatures the value of tan δ increases with rise in temperature, which may be due to space charge polarization, and interfacial polarization across the Ag/PZTP30 interface. The substitution of Pd in PZT shifts the phase transition towards lower temperature. The observed $T_c$ was further verified by differential scanning calorimetric (DSC) measurements as shown in Figure 3(b). The DSC curve indicates a sharp exothermic peak around 552 K (+/- 5K) corresponding to ferroelectric-paraelectric phase transition temperature. The temperature for an exothermic peak in DSC thermogram is nearly same as the ferroelectric phase transition temperature obtained from dielectric studies within the experimental uncertainty.

Micro Raman spectroscopy is a nondestructive characterization technique to shed light on the crystal structure, lattice disorder, vacancies, and phase transition temperatures in the material. Figure S6 depicts the Raman spectra of PZTP30 at three different temperatures whose Raman modes matched well with the PZT bulk and single crystal.[49,50,51] The tetragonal phase of bulk PZTP30 has been identified through Raman spectra, a significant shift in Raman modes towards the lower wavenumber for most of the modes has been observed with increase of temperature, which is either due to the softening of lattice with rise of temperature or the introduction of thermal disorder in this material. The change in bond length between oxygen and other cations will also decrease vibrational frequencies with increasing temperature. However, most of the Raman modes disappeared above the phase transition temperature, as can be seen in Raman spectra at 823 K. Hence the phase transition is further confirmed by a Raman spectroscopic study which was observed in temperature-dependent dielectric studies. The detailed analyses of temperature dependent Raman studies of this material will be reported separately.

For the conclusive evidence of existence of ferroelectricity at room temperature, electrical polarization (P) versus electric field (E) hysteresis loop measurements have been carried out on PZTP30 samples at room temperature and shown in Figure 4(a). The P-E hysteresis loop measurements were performed on a poled ceramic sample. The coercive field, remanent polarization and saturation polarization are found to be 7.7 kV/cm, 13.7 $\mu C/cm^2$, and 38.3 $\mu C/cm^2$ respectively with the maximum applied electric field (17 kV/cm). The presence of a ferroelectric hysteresis loop suggests the presence of ferroelectric properties in Pd substituted PZT ceramics. The well- saturated P-E loop of LSAT/LSMO/PZTP30 thin films has been also presented in the inset of Figure 4(a); its magnitude is higher than the bulk counterpart. The film was less conductive.

In order to prove the presence of magnetism, and to understand the origin of magnetism in PZTP30, magnetization as a function of magnetic field at room temperature is depicted in Figure 4(b). Figures S7(a-d) show the M(H) hysteresis behavior performed at various temperatures. Standard PZT pellets were synthesized in same conditions as standards and do not show any magnetic ordering, whereas the Pd-doped PZT shows room-temperature well saturated M-H curves with a large tail due to diamagnetic properties for higher applied magnetic field. Temperature-dependence of coercive field ($H_c$) and remanent magnetization ($M_r$) of PZTP30 is shown in lower inset of Figure 4(b), which monotonically increases with decrease in temperature. Since no impurity phase has been detected in XRD patterns of PZTP30, the observed room-temperature magnetism could be due to the presence of $Pd^{2+}/Pd^{4+}$ ions into the host lattice, leading to the emergence of ferromagnetic long-range ordering. The room-temperature ferromagnetism in some perovskite oxides have already been reported by substitution of ferromagnetic particles such as Ni, Fe, and Co into the host lattice.[52,53] The presence of $Pd^{2+}$ and $Pd^{4+}$ states in PZTP30 is viewed by us as the origin of magnetism, and the existence of $Pd^{2+}$ and $Pd^{4+}$ states has been already confirmed from the XPS and XRF studies (Figures 2 and S3). The zero-field-cooled (ZFC) and field-cooled (FC) behavior of PZTP30 at 500 and 1000 Oe from 25 to 300 K has been plotted in Figures S7((e) and 4(c) respectively**.** ZFC/FC data confirmed that there is no sharp transition up to 300 K. The results above prove that both ferroelectric and ferromagnetic ordering exists above room temperature and hence room temperature multiferroism.

For further understanding of the multiferroic nature, and the coupling between electric and magnetic order parameters in PZTP30 was investigated by studying the sample response to applied magnetic fields. The low-frequency magnetoelectric (ME) voltage coefficients ($\alpha_{ME}$) were measured. The sample was first poled at room temperature in an electric field of 30 kV/cm for 4 hours. The ME measurement system consisted of an electromagnet for applying a bias magnetic field H, a pair of Helmholtz coils for applying an ac magnetic field δH, and lock-in detection for measuring the ME voltage δV generated across the sample thickness. The ME voltage was measured as a function of H for H = 0-3 kOe, an ac field δH = 1 Oe at 100 Hz and at room temperature. The measurements were performed for two field orientations: first in-plane mode for H and δH parallel to each other and to the sample plane (along direction *1*) and perpendicular to δE along direction *3* (termed transverse orientation); and second, out-of-plane mode for all the three fields (H, δH, and δE) parallel to each other and perpendicular to sample plane (all the fields along direction *3* and termed longitudinal orientation).[42] The ME voltage coefficient $\alpha = \delta V/(t\ \delta H)$ where t is the sample thickness was estimated. The H dependence of longitudinal ($\alpha_{33}$) and transverse ($\alpha_{31}$) coupling coefficient for PZTP30 is plotted in Figure 5.

Figure 5 shows representative data on bias magnetic field (H) dependence of transverse and longitudinal ME voltage coefficients. Consider first the results for the longitudinal field direction ($\alpha_{E,33}$). As H is increased, $\alpha_{E,33}$ remains small for H < 0.7 kOe and then there is an increase in $\alpha_{E,33}$ with H to a maximum value of 0.36 mV/cm Oe at $H_m$= 3 kOe. Upon reversal the direction of H, there is a sign reversal in $\alpha_{E,33}$ (a phase shift of 180°). The magnitude of $\alpha_{E,33}$ is small compared to values for positive fields and reaches a maximum of 0.18 mV/cm Oe which is only 50% of the value for +H. This asymmetry in ME coefficient could be due to a magnetic anisotropy in the sample. The ME voltage coefficient vs H data do not show a peak or decrease to zero value for very high H due to saturation of magnetostriction. Similar observations were reported for several multiferroic composites.[54,55,56,57]

Figure 5 also shows α vs. H data for field perpendicular to the sample plane. The ME coefficient $\alpha_{E,31}$ shows a sign reversal relative to $\alpha_{E,33}$ and its magnitude increases almost linearly with H to a value of 0.15 mV/cm Oe. Upon reversal of direction of H, $\alpha_{E,31}$ becomes positive and shows a maximum value of 0.3 mV/cm.Oe, which is twice the magnitude for +H. Since the ME voltage arises due to magnetic-mechanical-electrical interactions, the voltage coefficients are directly proportional to the product of piezoelectric (*d*) and piezomagnetic (*q*) coupling factors.[55,56,57] Since the parameter $q=d\lambda/dH$, where $\lambda$ is the magnetostriction, the H-dependence of α is expected to tracks the slope of $\lambda$ vs. H. Saturation of λ at high field leads to $\alpha_E = 0$. In this particular system, however, the data in Fig.5 clearly indicate that saturation of λ does not happen for H = 3 kOe. For most ferromagnets, the longitudinal ($\lambda_L$) and transversal ($\lambda_T$) magnetostrictions follow the relation $\lambda_L=2\ \lambda_T$ and one expects $\alpha_{33}=2\ \alpha_{E,31}$. For the data in Fig.5, one estimates for H = 3 kOe $\alpha_{33}$ ~ 2 $\alpha_{E,31}$ for +H and $\alpha_{31}$ ~ 2 $\alpha_{E,33}$ for –H.

In the case of a bulk composite in which the magnetostrictive and the piezoelectric phases are uniformly mixed together, the transverse ME signal is formed by transverse magnetostriction whereas the longitudinal ME signal is due to the longitudinal magnetostriction. For a majority of ferromagnets the longitudinal magnetostriction is a factor of two higher than the transverse magnetostriction, and one expects $\alpha_L = 2\ \alpha_T$. Such an empirical relationship is confirmed in samples in the shape of cubes.[58] To measure both coefficients and to avoid the influence of demagnetizing fields, it is necessary to use long cylindrical samples. But such approach also leads to other difficulties: high voltage required for polarization and mismatch of input impedance of the measuring device with the impedance of the sample. One possible solution is to measure $\alpha_{E,T}$ for a disk sample of a bulk composite and then estimate the longitudinal coefficient from the empirical relationship.

**Conclusions:**

A novel single phase-pure PZTP30 magnetoelectric having tetragonal crystal structure with *P4mm* symmetry was discovered for possible room temperature multi-states tunable logic and nonvolatile memory elements under external E and M-fields. It possesses high magneto-electric coefficients ~ 0.36 mV/cm.Oe at $H_m$= 3 kOe in a single-phase system suggesting a strong coupling between piezo- and magneto-striction at nanoscale. It displays room-temperature weak ferromagnetism, strong ferroelectricity, and strong ME coupling. We believe the origin of magnetism is due to mixed valance states of the $Pd^{2+}/Pd^{4+}$ in PZT matrix as confirmed by XPS and XRF studies. A sharp ferroelectric-paraelectric phase transition is observed near 569 K, well supported by dielectric, Raman, and thermal studies. The giant room-temperature magneto-electric coupling makes it a future alternative of $BiFeO_3$ with a strong possibility for real device applications.

**Figure Captions:**

**Figure 1 Structural Characterization of PZTP30 Ceramic.** (a) The Rietveld refined XRD

patterns of PZTP30 ceramics using Fullprof Suite Software (b) The three-dimensional schematic sketch of the PZTP30 unit cell with tetragonal structure at room temperature.

**Figure 2 High resolution XPS spectra for evidence of existence of Pd.** (a) XPS spectra of Pd 3*d* deconvoluted into two peaks, the binding energies at 336.37 eV, 342.9 eV, and 337.53 eV, 343.43 eV are assigned to $Pd^{2+}$ and $Pd^{4+}$ respectively, (b) XPS spectra of O in PZTP30 ceramics at room temperature.

**Figure 3 Ferroelectric Phase Transition above room temperature** (a). Temperature dependence of relative dielectric constant of PZTP30 ceramics at different frequencies, the inset shows respective temperature dependent of $\tan\delta$, (b) DSC therogram of PZTP30 ceramics.

**Figure 4 Ferroelectric and Magnetic Order Parameters in PZTP30 Bulk** (a) Ferroelectric (P-E) hysteresis loops of PZTP30 at room temperature; the inset shows ferroelectric (P-E) hysteresis loops of LSAT/LSMO/PZTP30 thin films at room temperature, (b) Magnetic (M-H) hysteresis loops of PZTP30 at room temperature, M-H hysteresis loops of PZTP30 at 25 K (upper) and temperature dependence of $H_c$ and $M_r$ (lower) are in insets, (c) ZFC and FC plot of PZTP30 ceramics at 1000 Oe.

**Figure 5 Magnetoelectric coupling voltage coefficients $\alpha_{ME}$ of PZTP30 as a function of an externally applied magnetic field H.** Magnetoelectric coupling coefficients ($\alpha_{E33}$ and $\alpha_{E31}$) of PZTP30 ceramics at room temperature.

**Acknowledgement**

This work had financial support from the DOE-EPSCOR Grant # FG02-08ER46526. S. K., D. K. P. and K. P. would like to acknowledge fellowships from the NSF Grant # EPS – 01002410. We are also grateful to Dr. Satyaprakash Sahoo for his valuable suggestions in Raman studies.

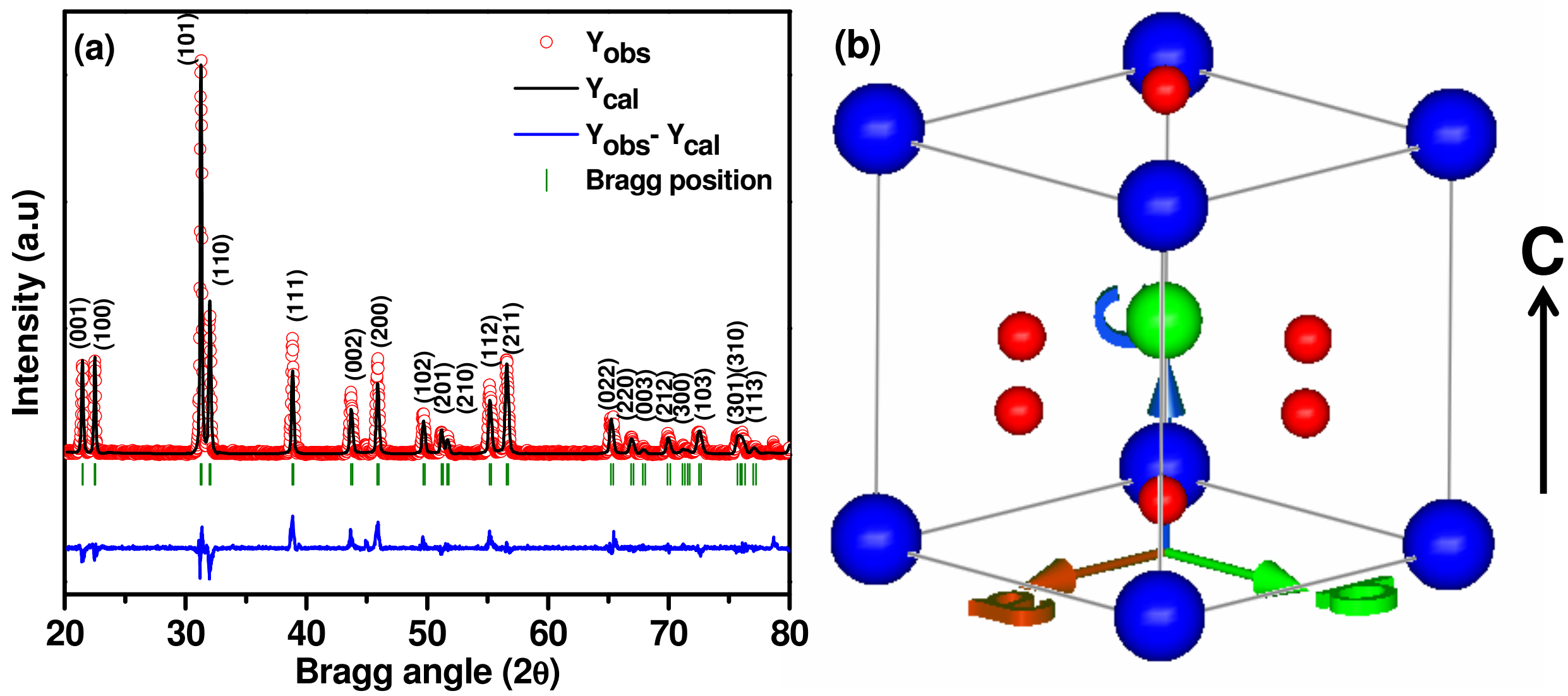

**Figure 1. Kumari et al.**

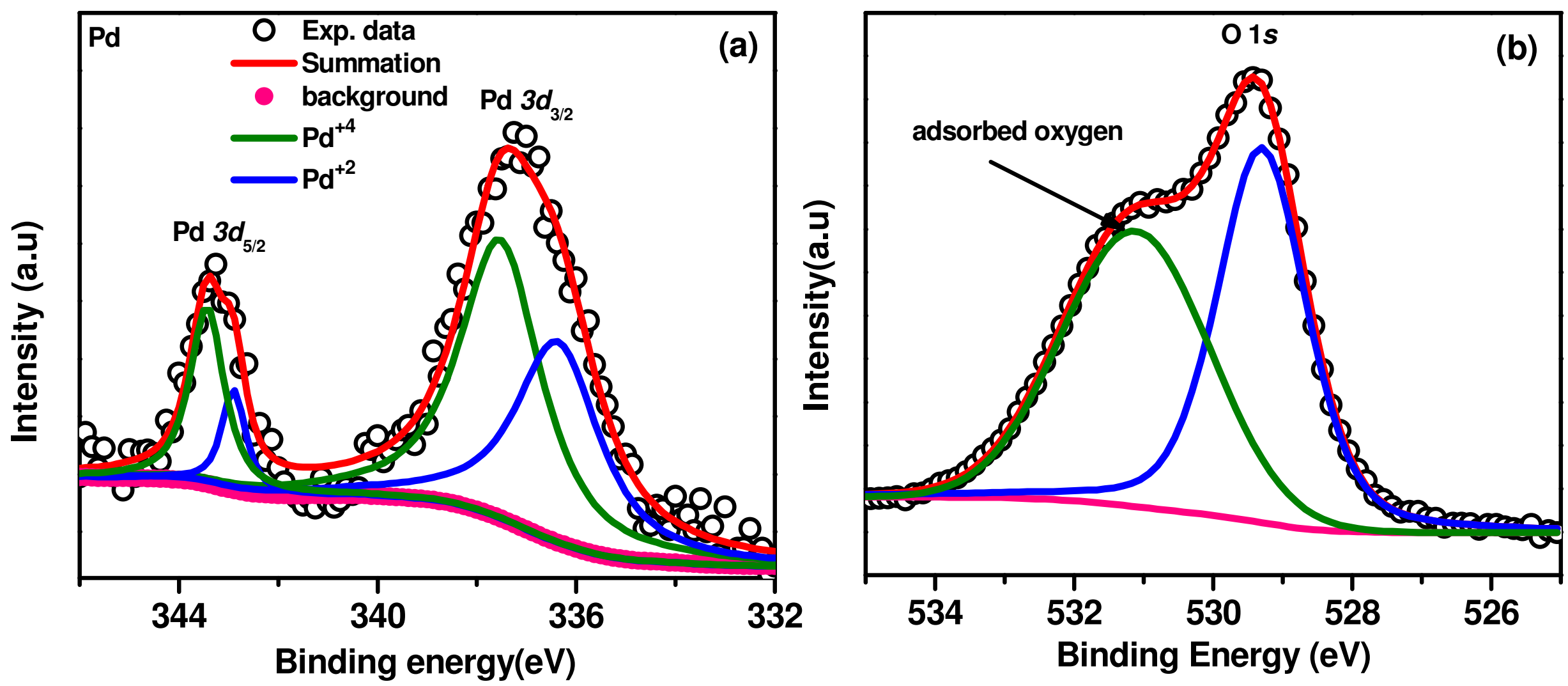

**Figure 2. Kumari et al.**

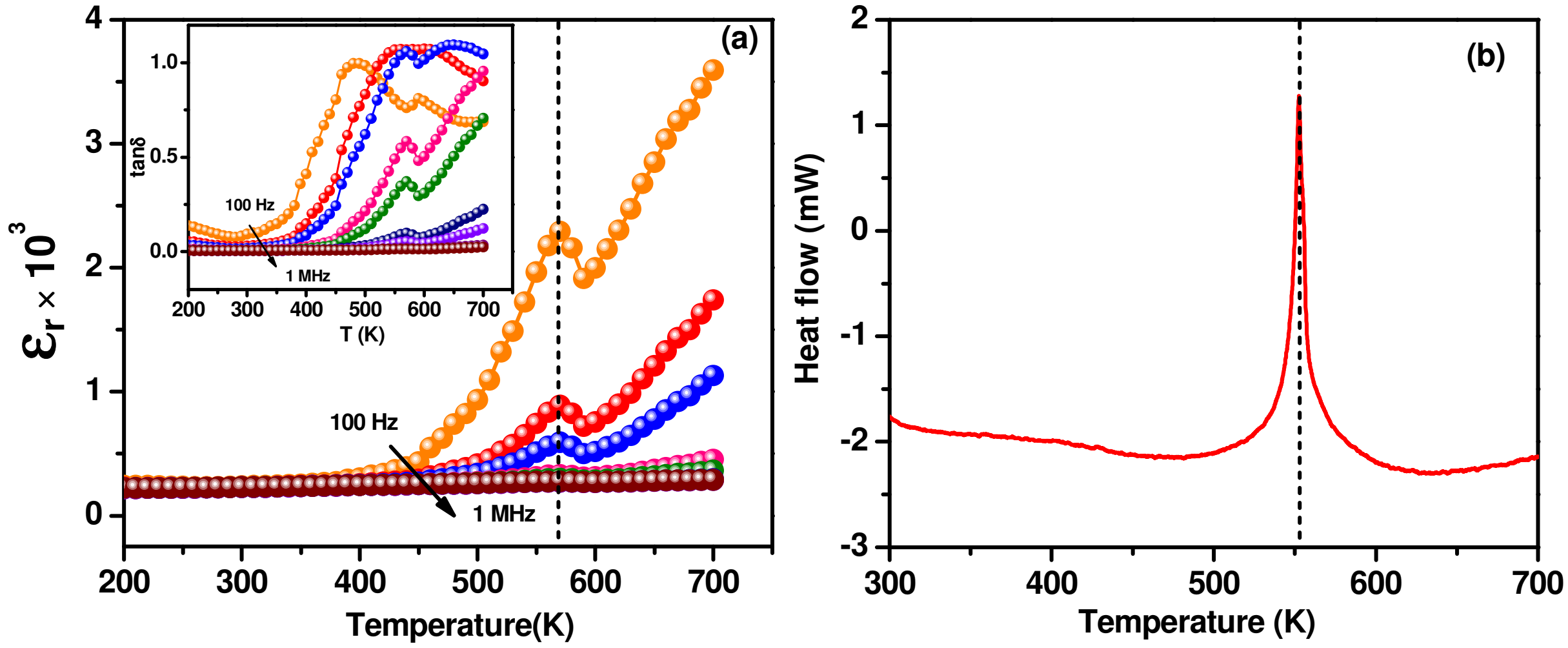

**Figure 3. Kumari et al.**

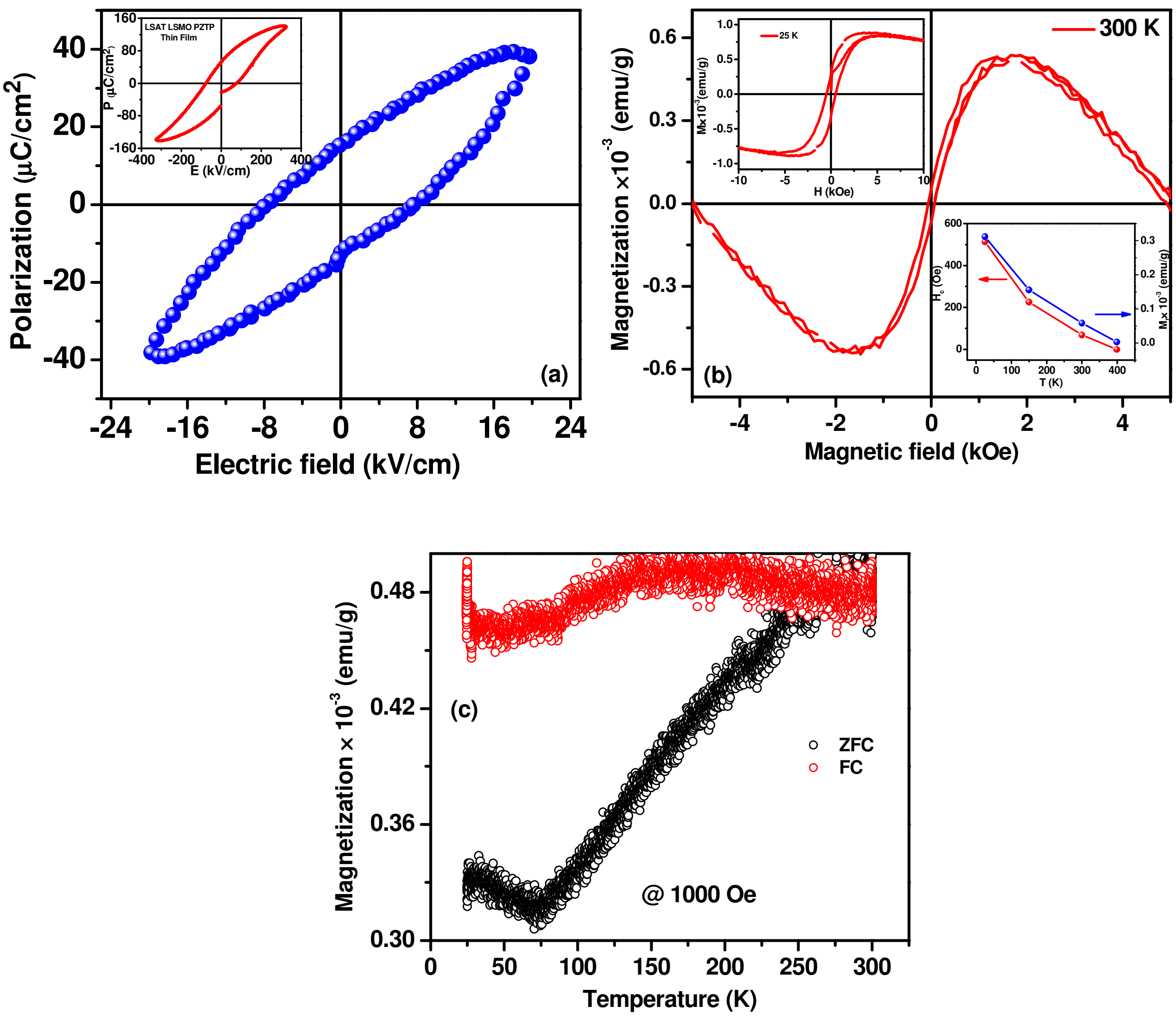

# Figure 4. Kumari et al.

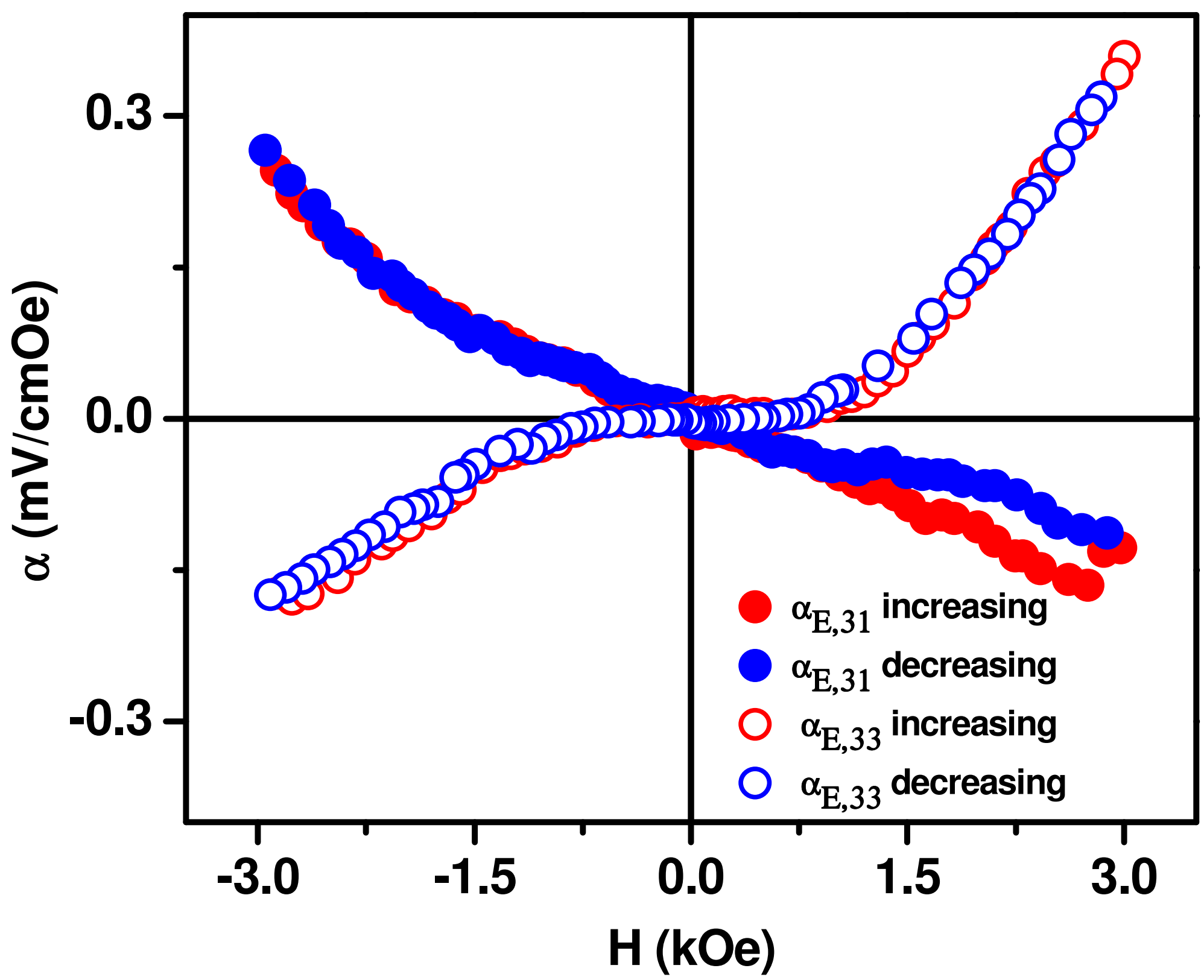

**Figure 5. Kumari et al.**

**Giant Magnetoelectric coupling in Single Phase $Pb(Zr_{0.20}Ti_{0.80})_{0.70}Pd_{0.30}O_{3-\delta}$ Multiferroics**

Shalini Kumari[1], Dhiren K. Pradhan[1], Nora Ortega[1], Kallol Pradhan[1], Christopher DeVreugd[2], Gopalan Srinivasan[2], Ashok Kumar[3], J. F. Scott[4], and Ram S. Katiyar[1,*]

[1]Department of Physics and Institute for Functional Nanomaterials, University of Puerto Rico, San Juan, PR 00931-3334, USA.

[2]Physics Department, Oakland University, Rochester, Michigan 48309-4401, USA.

[3]National Physical Laboratory (CSIR), Delhi, India.

[4]Department of Chemistry and Department of Physics, University of St. Andrews, St. Andrews KY16 ST, United Kingdom.

[*]Author to whom correspondence to be addressed. Electronic mail: rkatiyar@hpcf.upr.edu ( Ram S. Katiyar).

**Supplementary material 1 (S1)**

The Rietveld refinement of the diffraction patterns was performed by considering the tetragonal *P4mm* symmetry. The crystal structure parameters and reliability factors obtained after XRD refinement are listed in Table S1.

**Table S1**

| | |
|---|---|
| Molecular formula | $Pb(Zr_{0.20}Ti_{0.80})_{0.70}Pd_{0.30}O_{3-\delta}$ |
| Difractometer | Rigaku Ultima III |
| CuKα radiation | λ=1.5405 Å |
| Scan mode | θ-2θ |
| 2θ range | 20-80° |
| Scan width-scan speed | 0.02, 1° $min^{-1}$ |
| Crystal Symmetry | Tetragonal |
| Space group | *P4mm* |
| Unit cell parameters | a =b= 3.9531 c= 4.1401 Å<br>α= β= γ= 90° |
| Volume | 64.6964 $Å^3$ |
| Density | 15.128 $g/cm^3$ |
| Profile function | Pseudo-Voigt |
| FWHM parameters (U, V and W) | 36.07062, -0.15941, -0.23550 |
| Preferred orientation parameters | 0.1627, 0.0000 |
| Pattern residual ($R_p$) | 24.5 |
| Weighted pattern residual ($R_{wp}$) | 30.9 |
| Expected residual ($R_{exp}$) | 20.1 |
| Bragg factor ($R_B$) | 13.3 |
| Structural factor ($R_F$) | 7.59 |
| $Chi^2$ | 1.67 |

**Supplementary material 2 (S2)**

The bond length obtained after XRD refinement has been compiled in Table S2.

**Table S2**

| Atoms | Bond | Length(Å) | Occupancy |
|---|---|---|---|
| Pb | Pb-O2 | 2.5972 | 1 |
| Zr/Ti/Pd | Zr/Ti/Pd-O1 | 1.8122 or 2.3279 | 0.14/0.56/0.3 |
| | Zr/Ti/ Pd-O2 | 2.0511 | |
| O1/O2 | O1-O2 | 2.3464 | 0.5/0.5 |

**Supplementary material 3 (S3)**

The presence of all elements in PZTP30 bulk was confirmed through XRF measurement. The Table S3 contained percentage (%) of the actual molar mass and molar mass after XRF analysis of all individual elements in PZTP30 bulk after XRF analysis. The average XRF data matched with the initial elemental compositions taken for this study within the experimental limitations as shown in Table S3.

**Table S3**

| **Elements** | **Actual Molar Mass (%)** | **Molar Mass from XRF (%)** |
|---|---|---|
| Pb | 69.33 | 55.3 |
| Zr | 5.36 | 4.66 |
| Ti | 13.90 | 10.0 |
| Pd | 11.40 | 7.75 |
| O | 40.26 | 22.3 |

**Supplementary material 4 (S4): Morphology characterization**

The overall microstructure of PZTP30 pellets showed well defined densely packed grains with average size ranging between ~ 3–10 $\mu$m surrounded by distinct grain boundaries as shown in Figure S4. The presence of different shapes and sizes grains with neck to neck compaction reveled that the grain growth process was almost completed during the sintering process.

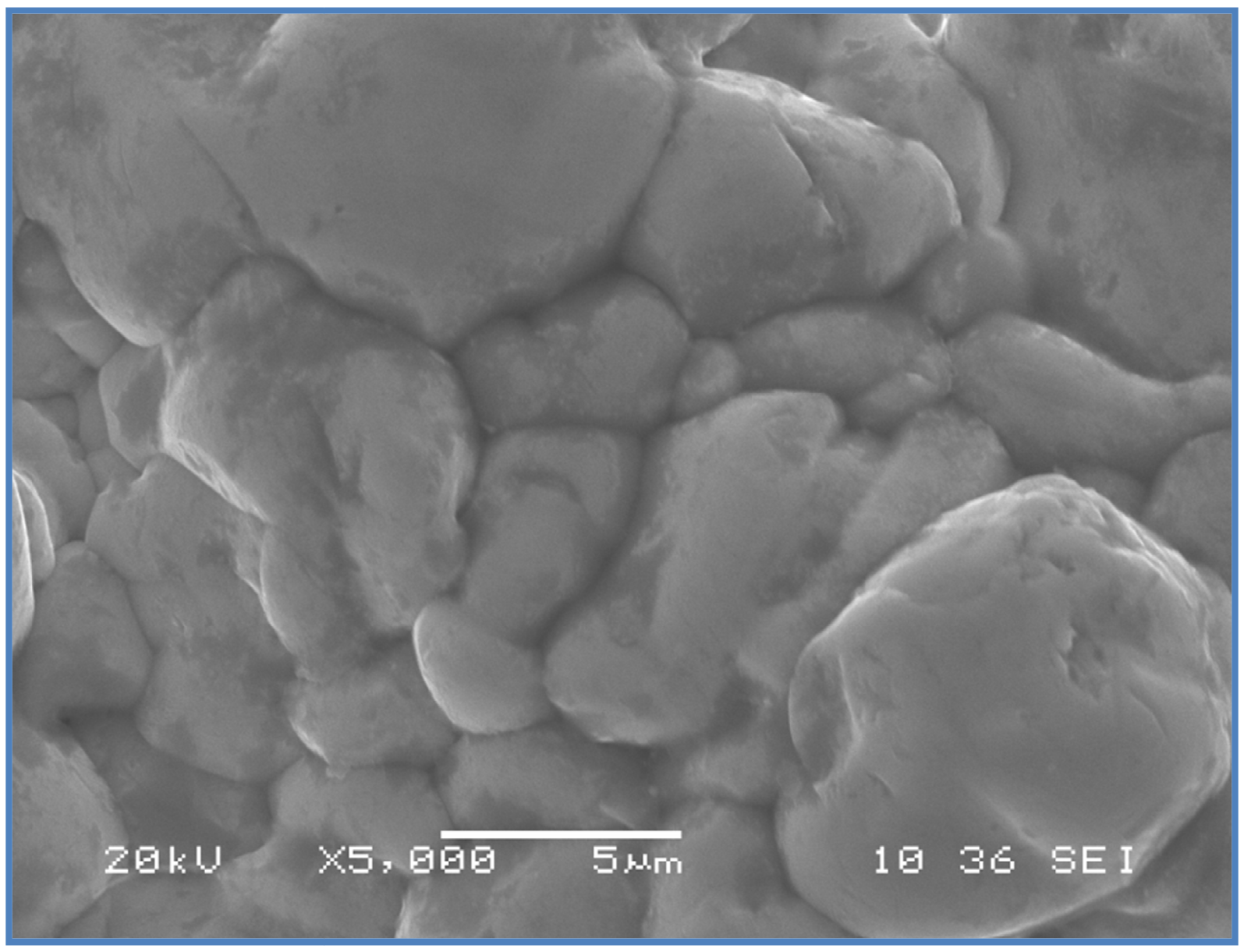

**Figure S4. Kumari et al.**

**Supplementary material 5 (S5): High Resolution XPS Spectra**

For the conclusive evidence of existence of all elements and their valance states, PZTP30 pellets were examined by high-resolution X-ray photoelectron spectroscopy (XPS). Figures S5 show the full XPS spectra of PZTP30 ceramics with the core level high resolution XPS spectra of Pb 4*f*, Zr 3*d*, Ti 2*p*. The peak positions in the XPS spectra were referenced to C 1*s* peak at 284.8 eV. The observed binding energy positions of Pb are Pb $4f_{7/2}$ =138.2 eV, and Pb $4f_{5/2}$ = 143 eV respectively. The high resolution XPS spectrum of Zr 3*d*, and Ti 2*p* splits up into two components due to spin-orbit effect. The spin-orbit doublet of Zr, and Ti were observed at following binding energies: Zr $3d_{5/2}$ = 181.3 eV, and Zr $3d_{3/2}$ = 183.7 eV; Ti $2p_{3/2}$ = 457.8 eV, and Ti $2p_{1/2}$ = 463.5 eV. These energy values confirm the valance states of $Zr^{4+}$, and $Ti^{4+}$ ions without any trace of $Ti^{3+}$ ions responsible for magnetism. The binding energies for all the individual elements were matched with the standard value. Moreover, the presence of all elements in PZTP30 bulk was confirmed clearly through XPS measurement.

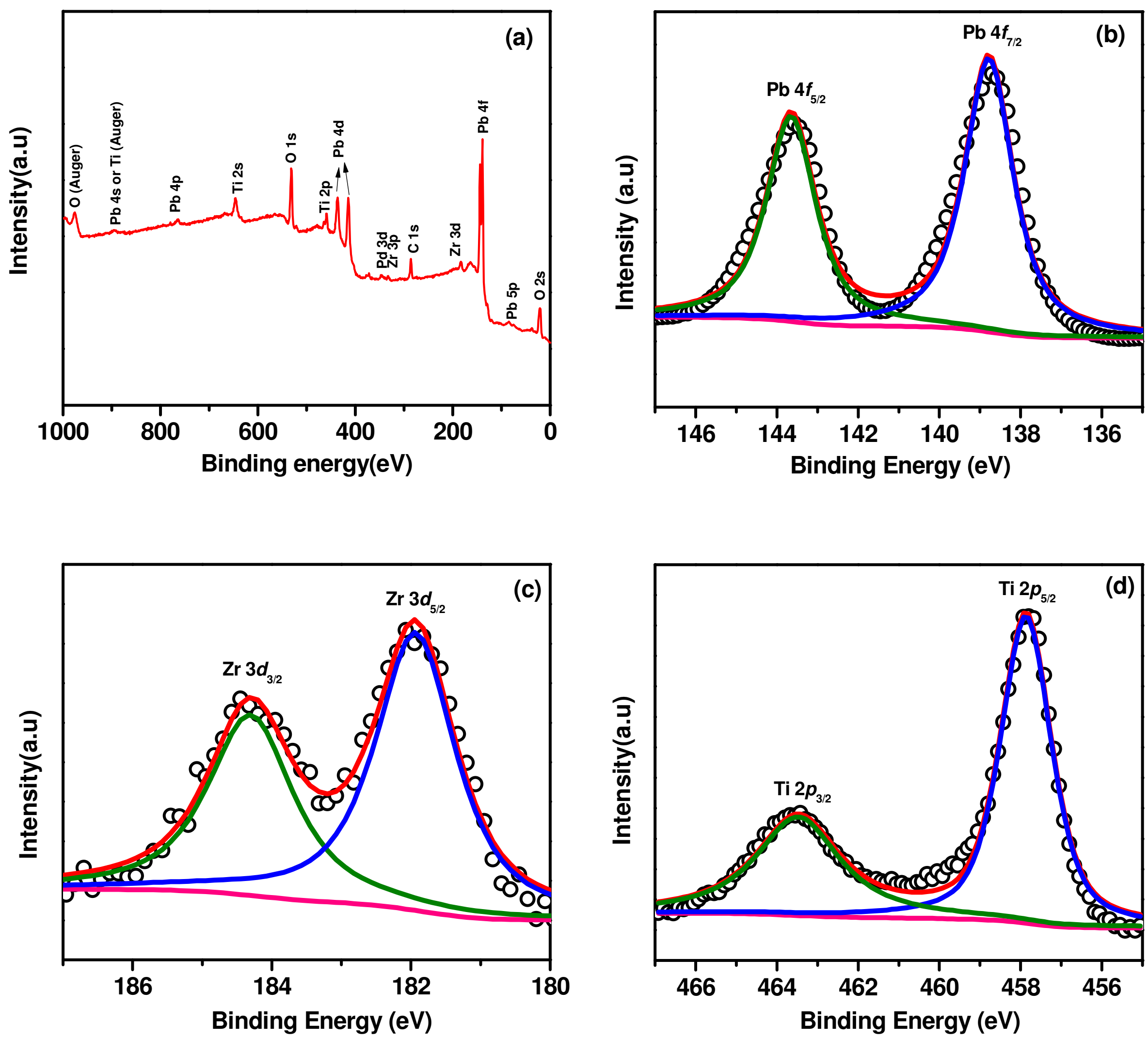

# Figure S5. Kumari et al.

**Supplementary material 6 (S6): Raman Spectroscopy of PZTP30 bulk**

Micro Raman spectroscopy is a nondestructive characterization technique to shed light on the crystal structure, lattice disorder, vacancies, and phase transition temperatures in the material. Figure S6 depicts the Raman spectra of PZTP30 at three different temperatures whose Raman modes matched well with the PZT bulk and single crystal. The tetragonal phase of PZTP30 bulk has been identified through Raman spectra, a significant shift in Raman modes towards the lower wavenumber for most of the Raman modes have been observed with increase of temperature, which is due to the softening of lattice with rise of temperature or the introduction of thermal disorder in this material. The change in bond length between oxygen and other cations will also decrease vibrational frequencies with increasing temperature. However, most of the Raman modes disappeared after phase transition as can be seen in Raman spectra at 823 K. Hence the phase transition is further confirmed by Raman spectroscopic studies which is already observed in temperature dependent dielectric studies. The detailed analyses of temperature dependent Raman studies of this material will be reported separately.

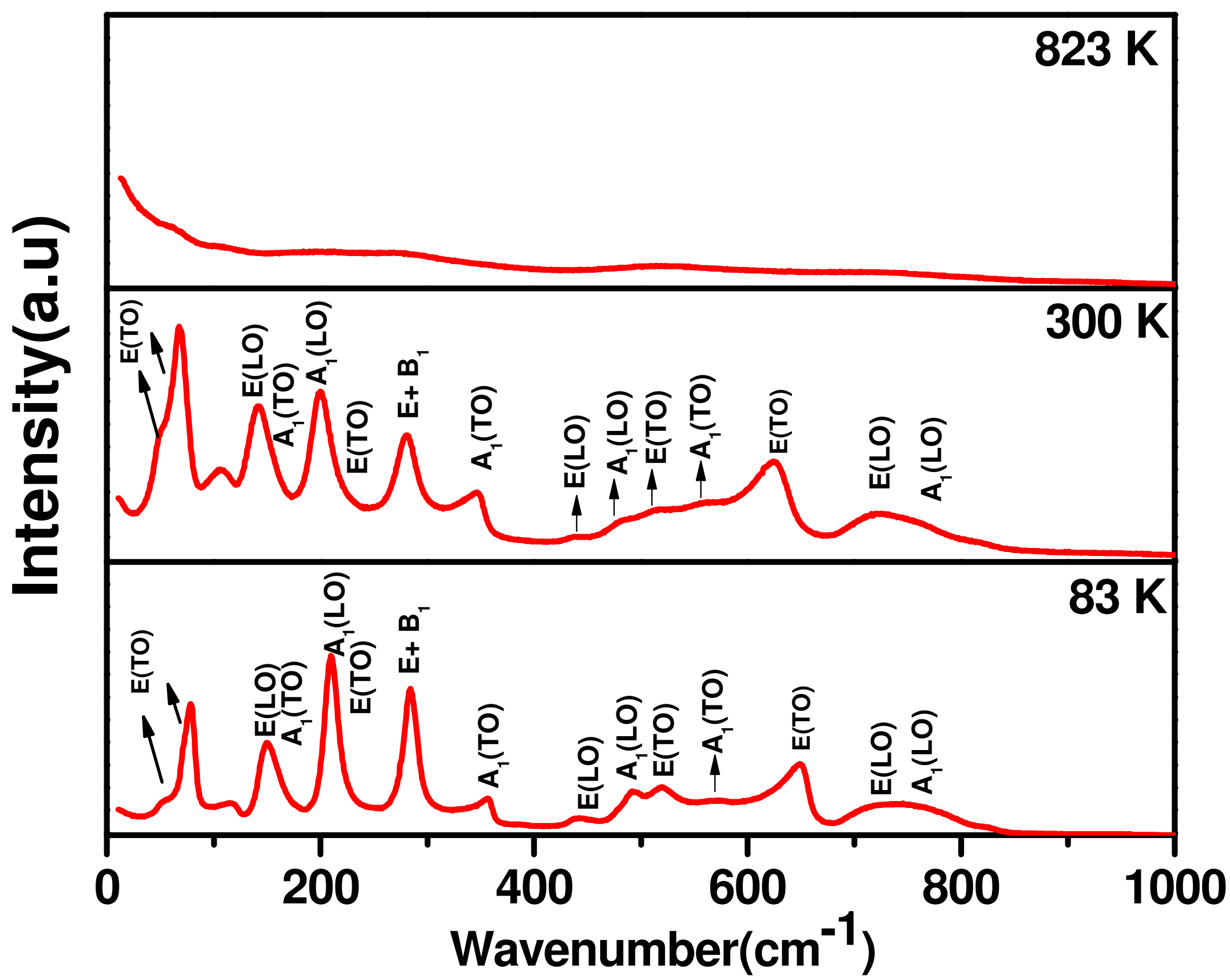

**Figure S6. Kumari et al.**

**Supplementary material 7 (S7): Magnetization of PZTP30 bulk.**

In order to prove the presence of magnetism and to understand the origin of occurrence of magnetism in PZTP30, magnetization as a function of magnetic field at various temperatures are depicted in Figure S7. The standard PZT pellets grown in same conditions do not show any magnetic ordering whereas the Pd doped PZT shows room temperature well saturated M-H curves with large tail of diamagnetic properties for higher applied magnetic field. Since no impurity phase has been detected in XRD patterns of PZTP30, the observed room-temperature magnetism could be due to the presence of $Pd^{2+}/Pd^{2+}$ ions into the host lattice lead to the emergence of ferromagnetic long-range ordering. The room-temperature ferromagnetism in some perovskite oxides have already been reported by substitution of ferromagnetic particles Ni, Fe and Co into the host lattice. The presence of $Pd^{4+}$ state in PZTP30 is the origin of magnetism and the existence of $Pd^{4+}$ state has been already confirmed from the XPS studies (Figure 2). The zero-field-cooled (ZFC) and field-cooled (FC) behavior of PZTP30 at 500 Oe from 25 to 300 K has been plotted in Figures S7((e)**.** ZFC/FC data confirmed that there is no sharp transition is found up to 300 K. The above results prove that both ferroelectric and ferromagnetic ordering exists above room temperature and hence room temperature multiferroism.

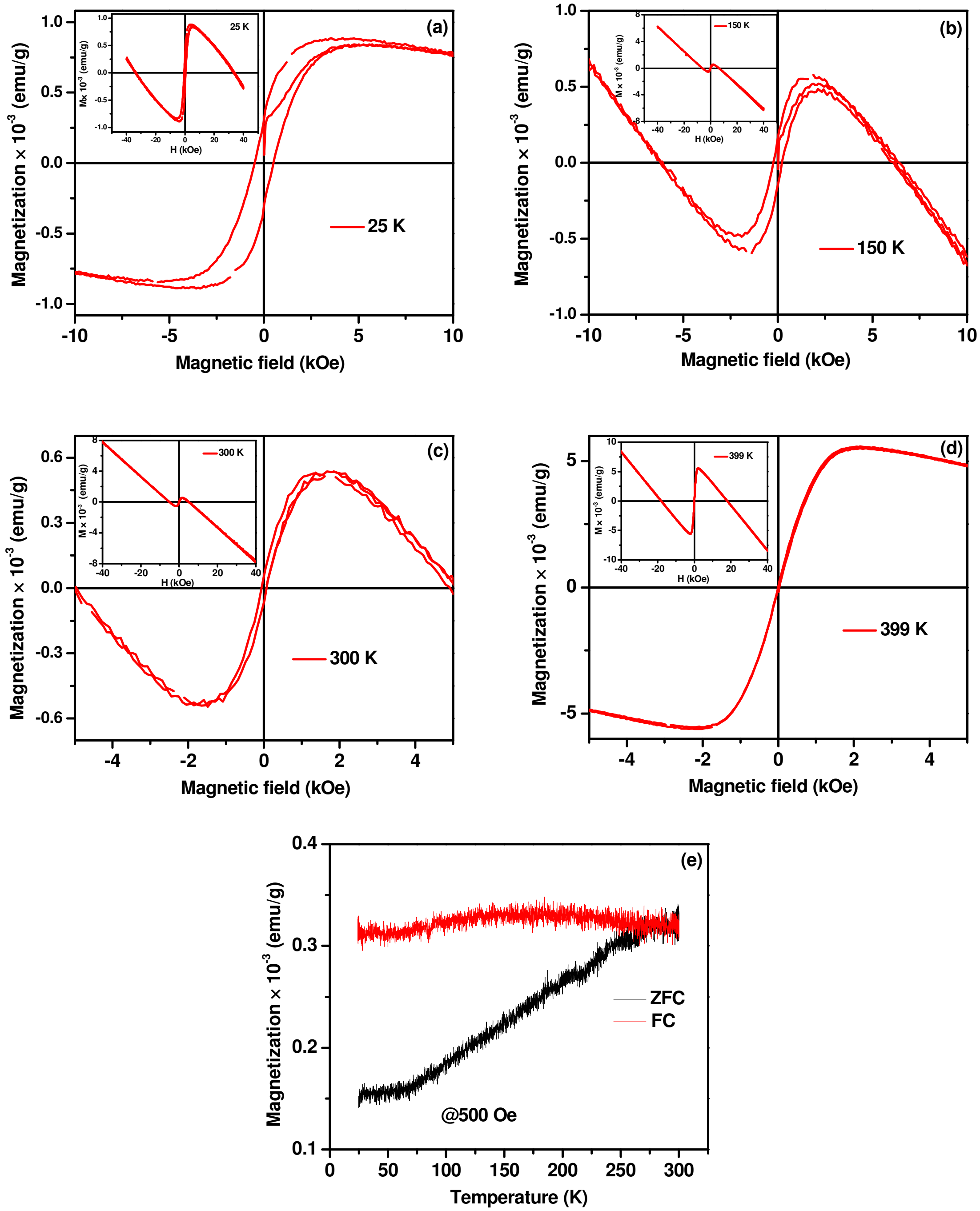

# Figure S7. Kumari et al.